%Paper: quant-ph/9412008
%From: mamwad@IREARN.BITNET
%Date: Wed, 28 Dec 1994 11:46:58 +0330

\baselineskip 1cm plus 0pt minus 0pt
\parskip 0pt plus 0pt minus 0pt
\font\vlbf=cmbx10 scaled 1440
\font\lbf=cmbx10 scaled 1200
\def\q#1#2{\eqno({\rm #1}.{#2})}
\hfill\vbox{\halign{&#\hfil\cr &quant-ph/9412008
                           \cr\noalign{\vskip -0.5cm} &IPM-94-073
                           \cr\noalign{\vskip -0.5cm} &TUDP 94-3
                           \cr\noalign{\vskip -0.5cm} &IASBS 94-7\cr}}

\vskip 1cm

\centerline{\vlbf A general formulation of discrete-time quantum mechanics,}

\centerline{\vlbf restrictions on the action and the relation of unitarity}

\centerline{\vlbf to the existence theorem for initial-value problems}

\vskip 1cm

\centerline{\bf M. Khorrami}

\centerline{\it Institue for Studies in Theoretical Physics and Mathematics}

\vskip -0.5cm\centerline{\it P. O. Box 19395-5746 Tehran, IRAN}
\vskip -0.5cm\centerline{{\it Fax: 98-21-2280415
E-mail: mamwad@irearn.bitnet}\footnote*{permanent address}}

\centerline{\it Department of Physics, Tehran University}

\vskip -0.5cm\centerline{\it North Kargar Ave., Tehran, IRAN}

\centerline{\it Institue for Advanced Studies in Basic Sciences}

\vskip -0.5cm\centerline{\it P. O. Box 45195-159, Gava Zang, Zanjan, IRAN}

\vskip 2cm

\centerline {\lbf Abstract }

\noindent A general formlulation for discrete-time quantum mechanics,
based on Feynman's method in ordinary quantum mechanics, is presented. It is
shown that the ambiguities present in ordinary quantum mechanics (due to
noncommutativity of the operators), are no longer present here. Then the
criteria for the unitarity of the evolution operator is examined. It is
shown that the unitarity of the evolution operator puts restrictions on the
form of the action, and also implies the existence of a solution for the
classical initial-value problem.
\vfil\break

\noindent{\lbf O~Introduction}

\noindent Discrete-space-time physics is an old tradition originated in
solid-state physics. This has been the starting point for lattice physics.
On the other hand, continuous-spce-time field theory has still some problems;
not only from the mathematical point of view, but also from the computational
point of view. The first kind of problems is, essentially, the lack of an
exact definition for functional integration, and the problem of ultraviolet
divergences in interacting field theories ([1], for example). The second
kind of problems is related to the fact that most of the numerical results
of field theory, are in fact perturbative results. A promising answer to
this kind of problems seem to be lattice~field~thories, specially
lattice~gauge~theories [2]. These theories, however, are interesting,
not only as an approximation for continuous-space-time theories, but also as
independent models [3-7].

Another problem, very much related to the above problems, is the lack of a
consistent theory of quantum gravity. In the context of quantum gravity,
there arises a natural scale for space-time, the plank scale, and it seems
plausible that something new should happen at this scale. In string theories,
this scale is the size, or the tension, of the string [8]. One can also use
this scale as the size of a possible space-time lattice. in fact, there is
no {\it true reason} for the continuousness of space-time: all of the
measurements of space-time (direct or indirect) have a certain resulotion,
which is very much larger than the plank scale. On the other hand, there are
examples for theories which force the time to be discrete ([9,10],
for example).

Also, there has been attempts to discretize the time in quantum mechanics
([11], for example). In this paper, first a general formulation of
discrete-time quantum mechanics is introduced (section I). This formulation
is based on the action principle, or the Feynman's path-integral formalism.

In section II the unitarity of the evolution operator is exploited to deduce
restrictions on the form of the action. This section addresses a very old
problem of classical mechanics: the problem of nonequivalent Lagrangians
which give rise to same equations of motion [12,13]. It is shown that in
one-dimensional space, all possible forms of the Lagrangian are equivalent
to the difference of a kinetic term and a potential term, where the potential
is an arbitrary function of the position. This means that, at least in
one dimension, any two Lagrangians (fulfilling the unitarity criteria and)
giving the same equation of motion, must be equivalent, up to a constant
multiplicative factor. At the end of this section, an example of an
{\it allowed} action is presented, which gives rise to the Lagrangian of a
charged particle moving in an electromagnetic field.

In section III, the equation of motion of the Hesinberg position operator is
investigated. It is shown that this equation is, unambiguously, deduced from
the classical equation of motion. In continuous-time quantum mechanics, this
is not, generally, the case; that is, one can not , in a straight forward
manner, deduce the equation of motion of the operators from the classical
equation of motion. At the end of this section, it is shown that the
unitarity criteria also guarantee the existence of a solution for the
classical initial-value problem.

\noindent{\lbf I~Formulation of discrete-time quantum mechanics
(Schr\"odinger picture)}

\noindent Ordinary (continuous-time) quantum mechanics is based on two
general parts: kinematics, which involves the definition of observables,
state space, and measurements; and dynamics, which discusses the evolution of
states. We need not change the kinematics. As for dynamics, we accept that a
quantum system is a linear dynamical system, the evolution of which is a
unitary one. In ordinary quantum mechanics, these among whith the principle
of correspondence lead to the Schr\"odinger equation:
$$\hat H\left\vert\psi\right>=i\hbar{d\over dt}\left\vert\psi\right>.\q I 1$$
In the case of discrete time, however, we do not have an infinitesimal
time-translation. So the Hamiltonian, which is the generator of
time-translation, does not arise naturally. One can of course write the
single-step evolution operator $\hat U$ as
$$\hat U(\tau )=\exp\left(-{i\tau\hat H\over\hbar}\right).\q I 2$$
But if $\hat H$ is a well-defined Hermitian operator, this is not a real
discretization of time; this is only a discrete sampling, because one can
also define $\hat U(t)$ for arbitrary $t$ as
$$\hat U(t)=\exp\left(-{i t \hat H\over\hbar}\right),\q I 3$$
and use it to find the solution of the Schr\"odinger equation (I.1).

There is an alternative which does not suffer from these artifacts, and that
is to use the Feynman path-integral formulation. One only needs to translate
the concept of path to discrete case, which is obvious. In ordinary quantum
mechanics we have
$$U(x',x'';t',t'')=A\int \left[Dx\left(t\right)\right]
\exp\left\{{i\over\hbar}S\left[x\left(t\right)\right]\right\} ;\q I 4$$
where $S$ is the action,
$$U(x',x'';t',t''):=\left<x'\right\vert\hat U(t',t'')\left\vert x''\right>,
\q I 5$$
and $\vert x'>$ is the eigenvector of the position operator. Integration is
over all paths which satiafy the boundary conditions
$$x(t')=x'\hbox{, and}\hskip 0.5cm x(t'')=x''\q I 6$$
This means that we are dealing with a ``multiple'' integral, whose measure
is (rather heuristically)
$$\left[Dx\left(t\right)\right]=\prod\limits_{t'>t>t''}dx(t).\q I 7$$

Although this description is rather ambiguous in continuous time, and must
somehow be regularized, it is completely clear in discrete time. In this case
we have
$$U_{n',n''}(x',x'')=A_{n',n''}\int\prod\limits_{n'>n>n''}dx_n
\exp\left[{i\over\hbar}S\left(x_{n'},\cdots ,x_{n''}\right)\right].\q I 8$$
Equivalently, one can use the single-step evolution operator
$$U_{n+1/2}(x',x''):=U_{n+1,n}(x',x'')=A_{n+1/2}
\exp\left[{i\over\hbar}S\left(x',x''\right)\right].\q I 9$$
We take this to be the axiom of time evolution, substituting (I.2). In (I.8)
and (I.9), $A$ is a constant independent of $x'$ and $x''$.

Now, as we are dealing with a dynamical system, we have
$$U_{n',n'''}(x',x''')=\int dx'' U_{n',n''}(x',x'')U_{n'',n'''}(x'',x'''),
\q I {10}$$
which leads to a familiar result for the action:
$$S_{n',n''}[x]+S_{n'',n'''}[x]=S_{n',n'''}[x].\q I {11}$$

To summerize, we define an action for a unit-time interval as
$$S_{n+1/2}(x_{n+1},x_n):=S_{n+1,n}(x_{n+1},x_n).\q I {12}$$
Then, we generalize this to arbitrary intervals as
$$S_{n',n''}(x_{n'},\cdots ,x_{n''}):=
\sum_{n'>n\geq n''}S_{n+1/2}(x_{n+1},x_n).\q I {13}$$
The single-step evolution operator is then
$$U(x,y):=A\exp\left[{i\over\hbar}S\left(x,y\right)\right].\q I {14}$$
We have droped the explicit time-dependence of the action for simplicity.
But, as one can readily see, this does not alter our later results.

Until now, we have not exploited the unitarity condition of $\hat U$, and
we have no restriction on the action. In the following section we will use
this condition to restrict the form of the action.

\noindent{\lbf II~Unitarity and restrictions on the action}

\noindent The unitarity condition for $\hat U$ is
$$\int dy U(x,y)U^*(z,y)=\delta (x-z),\q {II}1$$
or, in terms of the action,
$$\vert A\vert^2\int dy\exp\left\{ {i\over\hbar}\left[S\left(x,y\right)
-S\left(z,y\right)\right]\right\} =\delta (x-z).\q {II}2$$
Expanding the exponent as
$$S(x,y)-S(z,y)=(x-z)\cdot\nabla_x S(x,y)+O(\vert x-z\vert^2),\q {II}3$$
and defining
$$u:={1\over\hbar}(x-z),\q {II}4$$
it is easy to see that (II.2), to lowest order in $\hbar$, gives rise to
$$\vert A\vert^2\int dy\exp\left[{i\over\hbar}\left(x-z\right)\cdot\nabla_x
S\left(x,y\right)\right]=\delta (x-z).\q {II}5$$
The action, itself, may depend on $\hbar$. So, to be more exact, the above
equation holds for zeroth order (of $\hbar$) term of the action, i.e. the
classical action. But, as we are going to restrict the form of the classical
action, this does not mind.

Defining
$$Y:=\nabla_xS(x,y),\q {II}6$$
we have
$$\vert A\vert^2\int dy\left\vert{\rm det}\left(
{{\partial y}\over{\partial Y}}\right)\right\vert\exp\left[{i\over\hbar}
\left(x-z\right)\cdot Y\right]=\delta (x-z).\q{II}7$$
Now, the Jacobian is a function of $Y$ and $x$ only: It does not depend on
$x-z$. So the left-hand side of (II.7) is the Fourier transform of this
Jacobian, and is supposed to be the Dirac delta distribution. The Jacobian
should, therefore, be independent of $Y$. Since $A$ is independent of $x$,
the Jacobian should also be independent of $x$. We therefore have
$${\rm det}\left({{\partial y}\over{\partial Y}}\right)={\rm const.},
\q{II}8$$
or
$${\rm det}\left({{\partial Y}\over{\partial y}}\right)={\rm const.}
\q{II}8'$$

This is a nontrivial differential equation, which restricts the form of $S$.
If the dimension of the space is more than one, it is not easy to determine
the most general solution of this equation. But in one dimension the life is
simpler and one can obtain such a solution.

{\lbf II.I Action in one-dimensional space}

In one dimension, we have to solve
$${{\partial^2S}\over{\partial x\partial y}}=-{m\over\tau},\q{II}9$$
where we have written the constant as above for later convenience. $\tau$ is
the time step parameter. This equation can easily be solved:
$$S(x,y)=-{m\over\tau}xy+f(x)+g(y),\q{II}{10}$$
or
$$S(x,y)={m\over 2\tau}(x-y)^2-{\tau\over 2}\left[V\left(x\right)+
V\left(y\right)\right]+\left[\phi\left(x\right)-\phi\left(y\right)\right],
\q{II}{10}'$$
The third part of this expression has no effect on the dynamics of the
system. It is easy to see that addition of such a term to the action is
equivalent to the following gauge transformation
$$\vert x>\rightarrow\exp\left[-{i\over\hbar}\phi\left(x\right)\right]
\vert x>.\q{II}{11}$$
It is also easy to see that this term does not affect the equation of motion
(which we will encounter in section III). We therefore conclude that the
most general solution of (II.9) is equivalent to
$$S(x,y)={m\over 2\tau}(x-y)^2-{\tau\over 2}\left[V\left(x\right)+
V\left(y\right)\right]\q{II}{12}$$

This form is a very familiar one. In fact, if we divide the action by $\tau$
and let $\tau$ tend to zero, we obtain
$$L\left[x\left(t\right)\right]:=\lim_{\tau\rightarrow 0}{{S(x,y)}\over\tau}
={m\over 2}\dot x^2-V(x)\q{II}{13}$$
which is the standard form of one-dimensional Lagrangians used in textbooks.
This result is, however, an important one: First, in classical mechanics
the general Lagrangian formulation does not restrict the form of Lagrangian
as a function of position and velocity; one can have terms in Lagrangian
which are, for example, quartic in $\dot x$. However, nobody has ever needed
to consider such Lagrangians for real one-dimensional systems and it seems
that nature has chosen the special form (II.13). This formulation provides an
explanation for this fact, based one the general assumption of unitarity.
Second, there are many nonequivalent Lagrangians that lead to a same equation
of motion [12,13]. (By nonequivalent Lagrangians we mean Lagrangians, the
difference of them is not a total derivative.) One essentially has no way to
choose one of this Lagrangians for quantum theory. The discrete-time
formulation characterizes only one single Lagrangian among these.

{\lbf II.II Action in multi-dimensional space}

It is easily seen that an action like (II.12) satiafies $({\rm II}.8)'$.
To have a taste of other kinds of solutions, consider a perturbative
approach:
$$S(x,y)=S_0 (x,y)+s(x,y):={m\over 2\tau}(x-y)^2-{\tau\over 2}
\left[V\left(x\right)+V\left(y\right)\right]+s(x,y).\q {II}{14}$$
We want to solve the linearized equation, corresponding to $({\rm II}.8)'$,
in terms of $s$. This equation is readily seen to be
$${\rm tr}\left({{\partial^2s}\over\partial x\partial y}\right)=
\nabla_x\cdot\nabla_ys=0.\q{II}{15}$$
A special solution of this equation is of the form
$$s={1\over 2}\left\{\left[{\cal A}_1\left(x_1,y_2,y_3,\cdots\right)
-{\cal A}_1\left(y_1,x_2,x_3,\cdots\right)\right]
+\left[{\cal A}_2\left(y_1,x_2,x_3,\cdots\right)
-{\cal A}_2\left(x_1,y_2,x_3,\cdots\right)\right]+\cdots\right\}.\q{II}{16}$$
In the limit $\tau\rightarrow 0$, we have
$$s={\tau\over 2}\left\{\left[{{\partial{\cal A}_1}\over\partial x_1}\dot x_1
-\left({{\partial{\cal A}_1}\over\partial x_2}\dot x_2
+{{\partial{\cal A}_1}\over\partial x_3}\dot x_3+\cdots\right)\right]
+\left[{{\partial{\cal A}_2}\over\partial x_2}\dot x_2
-\left({{\partial{\cal A}_2}\over\partial x_1}\dot x_1
+{{\partial{\cal A}_2}\over\partial x_3}\dot x_3+\cdots\right)\right]
+\cdots\right\},\q{II}{17}$$
or
$$s=\tau\left({{\partial{\cal A}_1}\over\partial x_1}\dot x_1
+{{\partial{\cal A}_2}\over\partial x_2}\dot x_2+\cdots\right)-{\tau\over 2}
\left({{d{\cal A}_1}\over dt}+{{d{\cal A}_2}\over dt}+\cdots\right).
\q{II}{18}$$
The second term is a total derivative. Defining
$$qA_\alpha :={{\partial{\cal A}_\alpha}\over\partial x_\alpha},\q{II}{19}$$
we see that this solution corresponds to the Lagrangian
$$L={m\over 2}\dot x^2+q\dot x\cdot A-V(x),\q{II}{20}$$
which is the Lagrangian of a charged particle moving in the potential $V(x)$
and a magnetic field, the vector potential of which is $A(x)$.

\noindent{\lbf III Heisenberg picture, relation to classical mechanics, and
existence theorem for the solution of the equation of motion}

\noindent Evolution of the operators in Heisenberg picture is just like the
case of continuous time. We first define Heisenberg operators at time $n$
in terms of their Schr\"odinger counterparts as
$$\hat\Omega_n^H:=(\hat U^\dagger )^n \hat\Omega_n^S \hat U^n.\q{III}1$$
Now, we want to discuss the equation of motion for these operators; to be
more specific, we want to show that the Heisenberg position operator
satisfies the ``classical'' equation of motion. In fact, the true classical
equation of motion is (or can be) somehow ambiguous for operators, because
it involves the position operator at three distinct times, and these
operators do not commute with each other. But we will see that a certain
ordering is dictated from the evolution governed by (III.1).

Consider the matrix element $<x'\vert{\displaystyle
{{\partial S(\hat x^\uparrow ,w)}\over\partial w}}\vert x''>$, where we
define
$$\hat x^\uparrow :=\hat U^\dagger\hat x\hat U\hbox{, and}\hskip 0.5cm
\hat x^\downarrow :=\hat U\hat x\hat U^\dagger.\q{III}2$$
We have then
$$\eqalign{<x'\vert{{\partial S(\hat x^\uparrow ,w)}\over\partial w}\vert x''>
&=<x'\vert\hat U^\dagger{{\partial S(\hat x,w)}\over\partial w}\hat U
\vert x''>\cr &=\int dz<x'\vert\hat U^\dagger{{\partial S(\hat x,w)}
\over\partial w}\vert z>U(z,x'')\cr &=\int dz U^*(z,x'){{\partial S(z,w)}
\over\partial w}U(z,x'')\cr}.\q{III}3$$
Now, setting $w=x''$, we come to
$$<x'\vert{{\partial S(\hat x^\uparrow ,x'')}\over\partial x''}\vert x''>=
-i\hbar{\partial\over\partial x''}\delta (x'-x'')=:-<x'\vert\hat p\vert x''>,
\q{III}4$$
where the momentum operator $\hat p$ is the generator of space translation,
and its matrix element satisfies equation (III.4). To deduce (III.4) from
(III.3), we have exploited the form of $U$ in terms of the action, and also
unitarity of $\hat U$.

Equation (III.4) can be rewritten in terms of operators themselves:
$$<x'\vert{{\partial S(\hat x^\uparrow ,x'')}\over\partial x''}\vert x''>=
<x'\vert\overrightarrow\circ{{\partial S(\hat x^\uparrow ,\hat x)}
\over\partial \hat x}\overrightarrow\circ\vert x''>\hskip 1cm\Rightarrow$$
$$\overrightarrow\circ{{\partial S(\hat x^\uparrow ,\hat x)}\over
\partial \hat x}\overrightarrow\circ =-\hat p,\q{III}5$$
where the left-hand-side of (III.5) is an ordered form, in which $\hat x$s
are at the right of $\hat x^\uparrow$s.

By a similar arguement, or by taking the Hermitian conjugate of (III.5), one
concludes that
$$\overleftarrow\circ{{\partial S(\hat x^\uparrow ,\hat x)}\over
\partial \hat x}\overleftarrow\circ =-\hat p,\q{III}6$$
where the left-hand-side of (III.6) is in opposite ordering. Finally, since
in this case both of these orderings lead to the same result, we can define
a nonoriented ordering, which is equal to both of the above orderings:
$$\circ{{\partial S(\hat x^\uparrow ,\hat x)}\over\partial \hat x}\circ =
-\hat p.\q{III}7$$

By a similar arguement we come to
$$\circ{{\partial S(\hat x,\hat x^\downarrow )}\over\partial \hat x}\circ =
\hat p.\q{III}8$$
Now, eliminating $\hat p$ from (III.7) and (III.8), we obtain
$$\circ{{\partial S(\hat x,\hat x^\downarrow )}\over\partial \hat x}\circ +
\circ{{\partial S(\hat x^\uparrow ,\hat x)}\over\partial \hat x}\circ =0,
\q{III}9$$
or
$$\circ{{\partial S(\hat x_n,\hat x_{n-1})}\over\partial \hat x_n}\circ +
\circ{{\partial S(\hat x_{n+1},\hat x_n)}\over\partial \hat x_n}\circ =0.
\q{III}{10}$$
But this is the classical equation of motion for position operators
(obtained by extermizing the action), except that it is time ordered (either
in the forward, or in the reversed direction). It has no ambiguity and,
once we know the classical action, everything is determined.

Now we come to the problem of existence of solution in classical mechanics.
Suppose that we begin by an arbitrary action, which does not satisfy the
unitarity criterion. One can not guarantee that the classical equation of
motion has a solution for $x_{n+1}$ for any choice of $x_n$ and $x_{n-1}$:
this equation may be a nonlinear complicated one. This means that, not every
initial-value lead to a path: there may be a time when the particle can not
go anywhere.

Now suppose that the unitarity condition holds, so that the equation of
motion
for the operators, (III.7) through (III.9), hold. For any pair of initial
values $x_0$ and $x_{-1}$, one can solve the classical counterpart of (III.8)
for $p_0$. Use a Gaussian wave packet with
$$<\hat x>=x_0,\hskip 0.5cm <\hat p>=p_0,\q{III}{11}$$
and
$$\Delta x=\alpha\sqrt{\hbar\over 2},\hskip 0.5cm
\Delta p=\alpha^{-1}\sqrt{\hbar\over 2},\q{III}{12}$$
as an initial state for quantum mechanics. Then consider equation (III.9) in
the limit $\hbar\rightarrow 0$. In this limit, the uncertainities tend to
zero, the operators commute, and we have
$${{\partial S(<\hat x>,<\hat x^\downarrow >)}\over\partial <\hat x>}+
{{\partial S(<\hat x^\uparrow >,<\hat x>)}\over\partial <\hat x>}=0.
\q{III}{14}$$
Notice that there always exists a ``$<\hat x^\uparrow >$'', because one can
compute it through the Schr\"odinger picture.

Now, we know that
$${{\partial S(x_0,x_{-1})}\over\partial x_0}=p_0=<\hat p>=
{{\partial S(<\hat x>,<\hat x^\downarrow >)}\over\partial <\hat x>},
\q{III}{15}$$
where the last equality holds only in the limit $\hbar\rightarrow 0$.
Substituting $x_0$ and $x_{-1}$ for $<\hat x>$ and $<\hat x^\downarrow >$ in
(III.16), we will have
$${{\partial S(x_0,x_{-1})}\over\partial x_0}+
{{\partial S(<\hat x>,x_0)}\over\partial x_0}=0,\q{III}{16}$$
which means that $<\hat x^\uparrow >$ (in the limit $\hbar\rightarrow 0$)
satisfies the classical equation of motion for the initial-values $x_0$ and
$x_{-1}$. Specially, if the dimension of space is one, we have seen that any
action which satisfies the unitarity condition is of the form
$({\rm II}.10)'$, or equivalently (II.12). The equation of motion for such an
action is linear in $x_{n+1}$ and the coefficient of $x_{n+1}$ is nonozero.
Therefore every initial-value problem has one and only one solution.

In general, we have proved that the unitarity of the evolution operator,
constructed through (I.14) from the action, is sufficient for the existence
of a solution for classical initial-value problems.

\noindent{\lbf Acknowledgement}

\noindent I would like to express my deep gratitude to prof. R. Mansouri for
very useful discussions and encouragement.
\vfil\break

\noindent{\lbf References}

\item{[1]} J. Collins; {\it Renormalisation}, (1984) Cambridge University
Press

\item{[2]} C. Rebbi ed.; {\it Lattice Gauge Theories and Monte Carlo
Simulations, chapter 2}, World Scientific publisher

\item{[3]} F. J. Wegner; J. Math. Phys. {\bf 12} (1971) 2559

\item{[4]} K. G. Wilson; Phys. Rev. {\bf D10} (1974) 2445

\item{[5]} J. B. Kogut; Rev. Mod. Phys. {\bf 51} (1979) 659

\item{[6]} M.Khorrami; {\it Exact solution for the most general
one-dimensional minimally coupled lattice gauge theories},
Int. J. Theo. Phys., in press

\item{[7]} M.Khorrami; {\it Phase transition in one-dimensional lattice gauge
theories}, hep-th/9412090

\item{[8]} M. B. Green, J. H. Schwartz, E. Witten; {\it Superstring theory},
(1987) Cambridge University Press

\item{[9]} G. 't Hooft; Class. Quantum Grav. {\bf 10} (1993) 1023

\item{[10]} A. P. Balachandran and L. Chandar; {\it Discrete time quantum
physics}, SU-4240-579

\item{[11]} R. Friedberg and T. D. Lee; Nucl. Phys. {\bf B225} (1983) 1

\item{[12]} H. Helmholtz; Z. Reine Angew. Math. {\bf 100} (1887) 137

\item{[13]} G. Morandi, C. Ferrario, G. Lo Vecchio, G. Marmo, and C. Rubano;
Phys. Rep. {\bf 188} (1990) 147
\end